\documentclass[%
 reprint,
 amsmath,amssymb,
 aps,
pra,
]{revtex4-1}

\usepackage{graphicx}
\usepackage{dcolumn}
\usepackage{bm}
\usepackage{braket}

\begin{document}

\preprint{APS/123-QED}

\title{Cavity optomechanical sensing in the nonlinear saturation limit}

\author{Usman A. Javid$^{1}$}
\author{Steven D. Rogers$^{3}$}%
\author{Austin Graf$^{1}$}%
\author{Qiang Lin$^{1,2}$}%
\email{qiang.lin@rochester.edu}
\affiliation{$^{1}$Institute Of Optics, University of Rochester, Rochester NY 14627, USA}
\affiliation{$^{2}$Department of Electrical and Computer Engineering, University of Rochester, Rochester NY 14627, USA}
\affiliation{$^{3}$Johns Hopkins University, Applied Physics Laboratory, Laurel, Maryland 20723, USA}

\date{\today}

\begin{abstract}
Photonic sensors based upon high-quality optical microcavities have found a wide variety of applications ranging from inertial sensing,  electro- and magnetometry to chemical and biological sensing. These sensors have a dynamic range limited by the linewidth of the cavity mode transducing the input. This dynamic range not only determines the range of the signal strength that can be detected, but also affects the resilience of the sensor to large deteriorating external perturbations and shocks in a practical environment. Unfortunately, there is a general trade-off between the detection sensitivity and the dynamic range, which undermines the performance of all microcavity-based sensors. Here we propose an approach to extend the dynamic range significantly beyond the cavity linewidth limit, in the nonlinear modulation regime, without degrading the detection sensitivity for weak signals. With a cavity optomechanical system, we experimentally demonstrate a dynamic range six times larger than the cavity linewidth, far beyond the conventional linear region of operation for such a sensor. The approach demonstrated here will help design microcavity-based sensors to achieve high detection sensitivity and a large dynamic range at the same time, a crucial property for their use in a practical environment.
\end{abstract}

\maketitle


\section{\label{sec:level1}Introduction}
In a sensor design problem, two parameters are of the utmost importance. These parameters are the sensitivity which determines the smallest change in a stimulus that can be detected, and the dynamic range which determines the limit to which the stimulus strength can yield useful information. The upper limit of measurement of a sensor comes due to nonlinearities in either the physics of the sensing process or the detection and readout electronics used after the sensor. Ideally a linear input-output relationship is desired and the region of linearity determines this dynamic range. In a typical sensing experiment, there is a trade-off between sensitivity and dynamic range and a compromise has to be made in favor of one depending on the application.

\begin{figure}
  \includegraphics[scale=1]{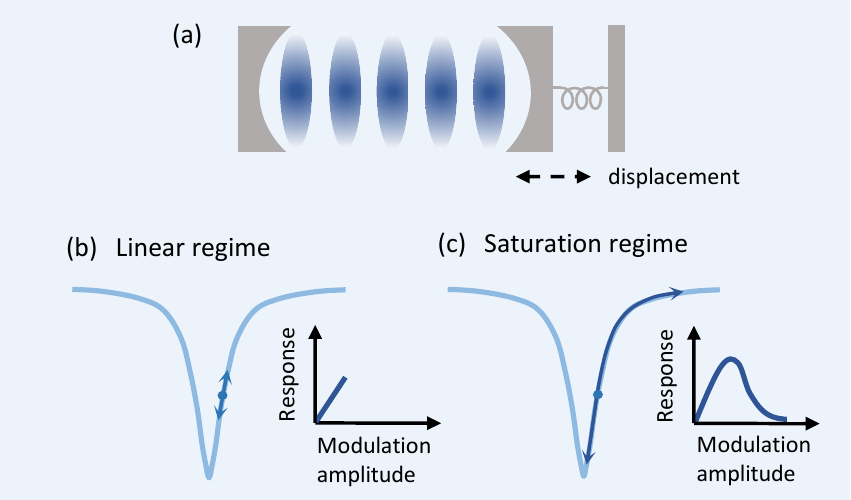}
  \centering
  \caption{Optomechanical modulation of a cavity: (a) An optical cavity with a movable mirror. The change in the length of the cavity shifts the resonance frequency. The response of an optical resonance to periodic change in cavity length yields (b) a linear response in the transmitted optical power when the modulation is small, and (c) nonlinear response when the modulation exceeds the cavity linewidth.}
  \label{fig1}
\end{figure}

An excellent example of this compromise is seen in optical sensors that use confined cavity modes \cite{rev_sensor4}. Such devices have been used in a variety of applications including particle sensing, biochemical analysis, inertial and force sensing, and electric and magnetic field sensing, to name a few \cite{rev_sensor4,bio1,rev_sensor5,rev_sensor1,rev_sensor6}. These sensors operate by measuring a change in the cavity’s resonance frequency due to a change in their environment. The detection sensitivity relies crucially on the optical quality of the sensing microcavity. Unfortunately, the nonlinear nature of the Lorentzian lineshape of the cavity resonance imposes an upper limit on the change in the resonance frequency that can be detected (see Fig. \ref{fig1}). Consequently, although a small cavity linewidth increases the detection sensitivity, it seriously limits the range of signal strength that can be probed. So far, the majority of research efforts on microcavity-based sensing focus on increasing the detection sensitivity \cite{rev_sensor4,bio1,rev_sensor5,rev_sensor1,rev_sensor6}. Little attention has been paid to address the limitation on the dynamic range, which is equally important for practical applications, as dynamic range also affects the robustness of the sensor against drifts and shocks in a changing environment. Here we propose and demonstrate an approach that can extend the dynamic range significantly beyond the cavity linewidth allowing measurement of strong signals, without impacting the detection sensitivity to measure weak perturbations. 

\par Depending on the application, a microcavity sensor can be operated in a static fashion to detect average signal strength, or in a dynamic fashion where the input signal carries a time-dependent modulation. The latter mode is particularly interesting since it is resilient to low-frequency noise. It has been widely employed in many areas, ranging from force \cite{force} and mass sensing \cite{mass1,mass2}, inertial sensing \cite{accel,accel2,gyro}, electro- \cite{E-field} and magnetometry \cite{magneto2}, acoustic sensing \cite{acoustic1,acoustic,acoustic2}, atomic force microscopy \cite{AFM}, refractive index sensing \cite{refrac},  and chemical and bio-sensing for tracking chemical reactions and molecular dynamics \cite{bio2,bio3,molecule}. Most sensing experiments make use of feedback control such as the Pound–Drever–Hall (PDH) technique \cite{PDH} to lock the laser to the cavity mode. The error signal of the feedback then gives the measurement. For static sensing, this extends the dynamic range limit from the cavity linewidth to the tuning range of the laser, potentially with the help of a frequency reference \cite{fiber-sensor1,fiber-sensor2}. However, for dynamic sensing, the bandwidth of the feedback electronics limits the dynamic range as well as the stimulus frequency that can be tracked, with the cavity linewidth still limiting the dynamic range for signals outside the bandwidth \cite{fiber-sensor3,fiber-sensor4}. This makes dynamic sensing problematic, dependent on electronics rather than the physics of the sensor which creates many complications. For example, the nonlinearities in the laser’s tuning response can make the measurement inaccurate, potentially requiring a frequency reference. Further complications occur in the resolved-sideband regime, where the modulation frequency exceeds the cavity linewidth and the cavity mode can no longer move with the stimulus. This brings a need for a different approach for sensing that is independent of these factors.

\par In this article, we propose and demonstrate an approach that can extend the dynamic range significantly beyond the cavity linewidth independent of the signal frequency. We show that, by simply detecting the first three harmonics of the transduced modulation signal, we will be able to extend the dynamic range of a cavity sensor, remarkably, to nearly an arbitrarily large value. We demonstrate this feature for both the adiabatic regime and the resolved-sideband regime where the modulation frequency is below and above the cavity linewidth, respectively. With a cavity optomechanical system, we experimentally measure modulation amplitudes over six times larger than the cavity linewidth, far beyond the linear region of operation for such a sensor. Our analysis shows that this approach of extending the detection bandwidth to increase the dynamic range will reduce detection sensitivity for large signals that would otherwise saturate the cavity. However, by combining our proposed technique with the traditional method of sensing at the same frequency as the stimulus, the high sensitivity required to measure weak perturbations can be preserved with the extended dynamic range. Our proposed approach will benefit applications in metrology that require versatile sensors with high sensitivity and large dynamic range capable of probing both strong and weak signals.

\begin{figure*}
\centering
  \includegraphics[scale=1]{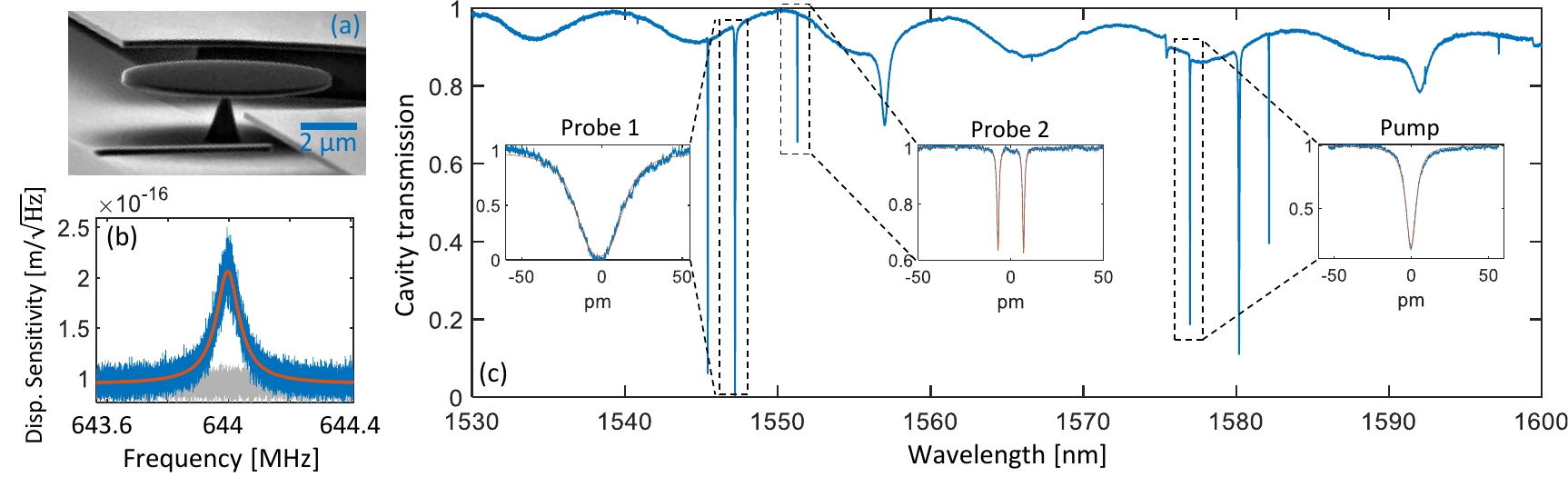}
  \caption{(a) SEM image of the silicon microdisk resonator. (b) Optically transduced thermal noise spectrum of the radial breathing mode of the microdisk centered at 644 MHz with Lorentzian fitting. The noise floor of the detector is shown in gray. (c) Transmission spectrum of the resonator obtained by scanning a tunable laser through the device. The three insets in the figure show detailed traces of the resonances used in phonon lasing experiments with loaded quality factors of $1.95\times 10^5$ for Pump, $5.7\times 10^4$ for Probe 1, and $1.23\times 10^6$ for Probe 2.}
  \label{fig2}
\end{figure*}

\section{Theory} \label{sectheory}
In a typical sensing experiment, as the modulation amplitude exceeds the cavity linewidth, the transduced energy tends to shift to higher harmonics of the modulation frequency due to the nonlinear transmission of the cavity resonance. It stands to reason that an ever-increasing detection bandwidth will be needed to sense increasingly large modulation signals. Earlier work on estimating temporal modulation of a cavity resonance has relied on using a Taylor series expansion to approximate the cavity transmission around a mean detuning \cite{taylor2,Taylor3}. This only works in the adiabatic regime, where the modulation frequency is much smaller than the cavity linewidth and the expansion diverges as the modulation amplitude exceeds this linewidth, after which numerical estimation of the cavity transmission \cite{taylor1} can be done. We circumvent this issue by leveraging the periodic nature of the transmitted optical power to determine higher order terms of a Fourier series from lower order terms, irrespective of the regime of operation.
\par Consider a cavity mode $a(t)$ with a resonance frequency $\omega_0$ and an intrinsic linewidth $\Gamma_0$. The cavity mode is excited with a laser with a field amplitude $A_{in}$ and frequency $\omega_L$ and the resonance frequency is modulated in time with an amplitude $G(t)$. In a frame rotated with the laser frequency, the equation of motion of the optical field inside the cavity is given by
\begin{align} \label{eqdiff}
    \frac{da(t)}{dt}=[i(\Delta-G(t))-\Gamma_t/2]a(t) +i\sqrt{\Gamma_{ex}}A_{in},
 \end{align}
where $\Delta=\omega_0-\omega_L$, $\Gamma_{ex}$ is the external coupling rate and $\Gamma_t=\Gamma_0+\Gamma_{ex}$. For an optomechanical system $G(t)=g_{om}x(t)$, where $g_{om}$ is the optomechanical coupling strength and $x(t)$ is the position of the moving cavity wall. For a sinusoidal modulation at a frequency $\Omega_m$, i.e. when $x(t)=x_0 \cos{(\Omega_m t)}$, Eq. \eqref{eqdiff} can be exactly solved in the steady state using a Floquet approach giving a solution \cite{diffeqn1,diffeqn2}
\begin{align} \label{eqexact}
    a(t)=-i\sqrt{\Gamma_{ex}}A_{in} \sum_{n,p}\frac{J_n(\eta)J_{n-p}(\eta)}{i(\Delta-n\Omega_m)-\Gamma_t/2}\exp[{ip\Omega_m t}], 
 \end{align}
where $J_n(\eta)$ is the Bessel function of the first kind and order $n$, and $\eta=g_{om}x_0/\Omega_m$ is the scaled modulation amplitude. This expression can be simplified by making some approximations. In the resolved-sideband regime, where $\Omega_m \gg \Gamma_t$, we can approximate
\begin{align} \label{eqresol}
    a(t) \approx -i\sqrt{\Gamma_{ex}}A_{in}J_0(\eta) \sum_{p}\frac{J_{-p}(\eta)}{i\Delta-\Gamma_t/2}\exp{[ip\Omega_m t]}, \nonumber \\
    =-\frac{i\sqrt{\Gamma_{ex}}A_{in}J_0(\eta)}{i\Delta-\Gamma_t/2}\exp[{-i\eta \sin{(\Omega_m t)}}].
 \end{align}
Under this approximation, the cavity field is only phase modulated. The power transmitted from the cavity $P(t)$ is then given by
\begin{align}
    P(t)=|A_{in}+i\sqrt{\Gamma_{ex}}a(t)|^2 .
 \end{align}
With the approximation made in Eq. \eqref{eqresol}, the time varying part of the transmitted power is calculated to be (See Appendix \ref{AppEqn} for details)
\begin{align} \label{eqapprox}
    \delta P(t)&=\frac{2|A_{in}|^2 J_0(\eta)\Gamma_{ex}}{\Gamma_t} [  J_0(\eta) + 2\sum^{\infty}_{n=1} J_{2n}(\eta)\cos{(2n\Omega_m t)} \nonumber  \\      &+2\sum^{\infty}_{n=0} J_{2n+1}(\eta)\sin{((2n+1)\Omega_m t)} ],
 \end{align}
where we have set $\Delta=\Gamma_t/2$. Equation \eqref{eqapprox} shows that in the resolved sideband regime, the harmonics of the modulation signal in the transmitted power are scaled by the Bessel function of the same order. Due to this, we can exploit the recurrence relation of the Bessel function
\begin{align} \label{eqbessel}
    2n J_n(\eta)=\eta[J_{n+1}(\eta)+J_{n-1}(\eta)].
 \end{align}
Therefore, using the first three harmonics, the modulation amplitude $x_0$ can be unambiguously given by
\begin{align} \label{eqamp}
    g_{om} x_0=4\Omega_m \frac{P_2}{P_1+P_3},
 \end{align}
where $P_n$ is the amplitude of the transmitted power at the $n$-th harmonic. In sensing experiments, usually the transmitted power at $\Omega_m$ is measured and it saturates as the modulation amplitude approaches the linewidth giving a maximum measurable amplitude of $g_{om}x_{max}\approx \Gamma_t/2$. We will refer to this as the saturation limit. Equation \eqref{eqamp} shows that the amplitude of the modulation signal can be measured even when it is large enough to saturate $P_1$ by using only the next two harmonics. Therefore, a finite bandwidth is sufficient for measuring arbitrarily large modulation.
\par The calculation shown here requires operation in the resolved-sideband regime for measuring modulation beyond the saturation limit of the cavity. However, it is important to note that most sensors operate in the adiabatic regime where $\Omega_m \ll \Gamma_t$. It is not immediately clear in Eq. \eqref{eqexact} how such an approximation can be made. Another approach to solve Eq. \eqref{eqdiff} is by integration. We can obtain the same solution as
\begin{align} 
    a(t)&=i\sqrt{\Gamma_{ex}}A_{in}\int^{\infty}_0 d\tau \exp{[(i\Delta-\Gamma_t/2)\tau]} \nonumber \\
    &\times \exp{ \left[ -i\int^{\tau}_0 d\tau' G(t-\tau')\right] }.
 \end{align}
In the adiabatic limit, $G(t)$ varies much slower than the cavity lifetime. Therefore, we can approximate
\begin{align} 
    a(t) \approx i\sqrt{\Gamma_{ex}}A_{in}\int^{\infty}_0 d\tau \exp{[(i\Delta-\Gamma_t/2)\tau]}\exp{\left[ -i \frac{G(t)}{\Gamma_t/2}\right] }.
 \end{align}
Setting $G(t)=g_{om}x_0 \sin{(\Omega_m t)}$ and integrating, we get
\begin{align} \label{eqadia}
    a(t) = \frac{-i\sqrt{\Gamma_{ex}}A_{in}}{i\Delta-\Gamma_t/2}\exp{[-i \frac{g_{om}x_0}{\Gamma_t/2}\sin{(\Omega_m t)}]}.
 \end{align}
Comparing Eq. \eqref{eqadia} to Eq. \eqref{eqresol}, we see that a similar relationship for $x_0$ as Eq. \eqref{eqamp} can be obtained as
\begin{align} \label{eqapprox2}
    g_{om} x_0=2\Gamma_t \frac{P_2}{P_1+P_3}.
 \end{align}
It is quite interesting to see that similar results can be obtained in two extreme conditions of modulation. In contrast, if a Taylor expansion is used, the modulation amplitude can be approximated as $P_2/P_1$ \cite{taylor1} only in the adiabatic limit by setting $\frac{da}{dt}=0$. It is important to note that when the amplitude exceeds the saturation limit, the adiabatic approximation breaks even when $\Omega_m \ll \Gamma_t$. In Section \ref{secdisc}, we will show that the proposed approach actually works far beyond this approximation.

\begin{figure*}[ht!]
  \includegraphics[scale=0.91]{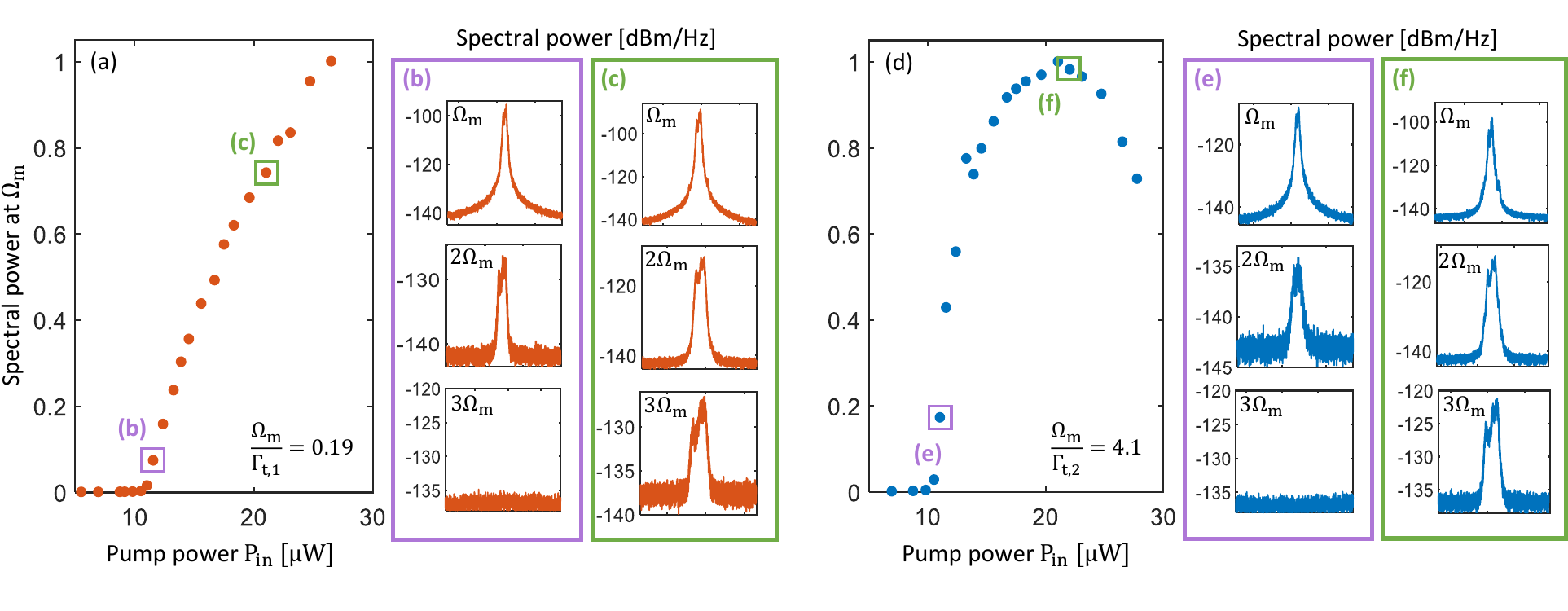}
  \centering
  \caption{(a) Integrated RF spectral power at $\Omega_m$ with increasing pump power transduced by Probe 1 showing phonon lasing behavior, driven by the pump mode, with a lasing threshold of 10.8 $\mu$W. (b) and (c) RF spectrum at the first three harmonics at two points identified in (a). (d) Integrated RF spectral power at $\Omega_m$ with increasing pump power for Probe 2 with (e) and (f) showing RF spectra at the three harmonics at two point identified in (d). The data in (a) and (d) is normalized to the corresponding maxima.}
  \label{fig3}
\end{figure*}

\begin{figure*}
  \includegraphics[scale=0.93]{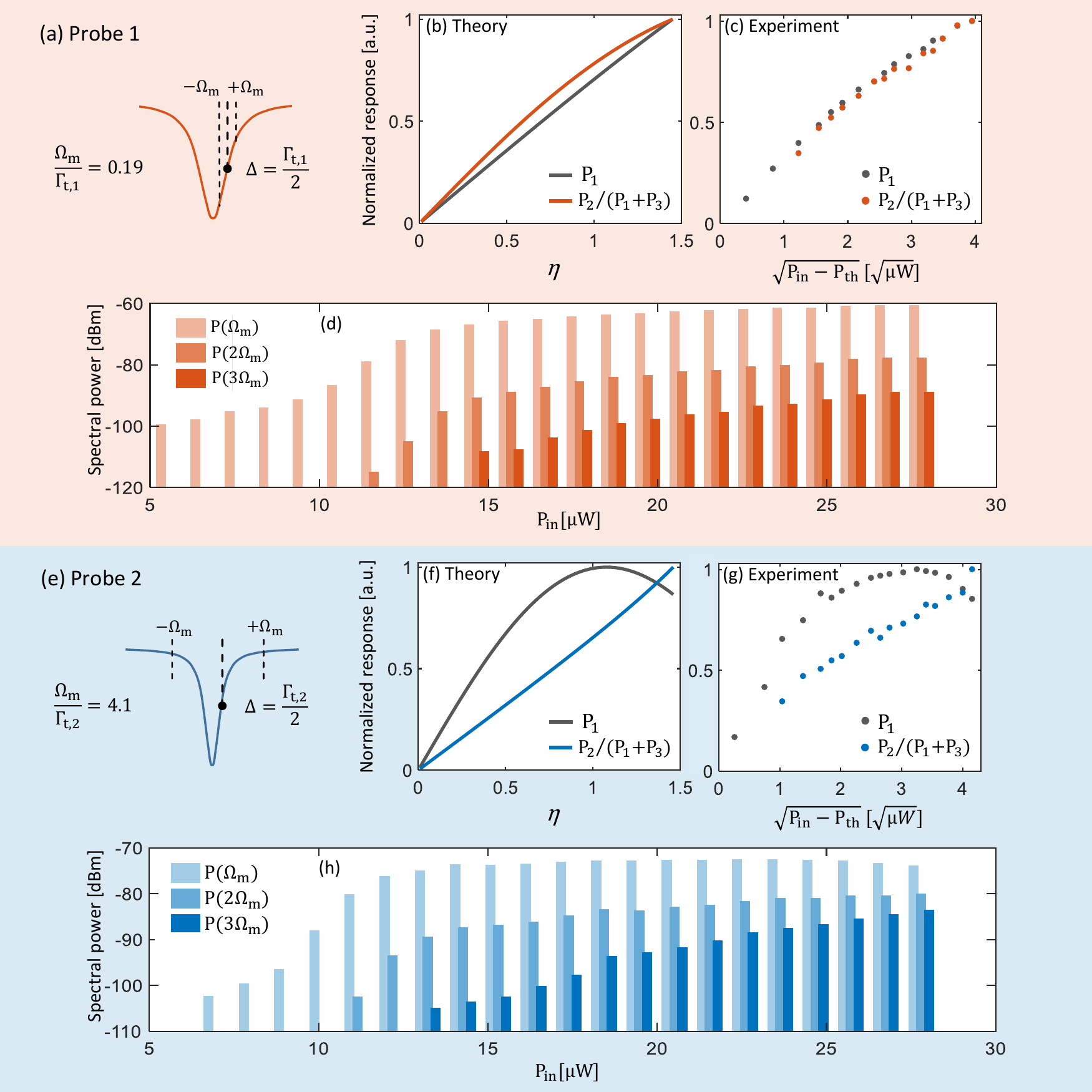}
  \centering
  \caption{Sensing mechanical motion beyond the saturation limit. (a) Probe 1 laser is locked to the resonance at a detuning of $\Delta=\Gamma_{t,1}/2$ with the mechanical sidebands forming well within the resonance putting it in the unresolved-sideband regime. The theory and experimentally obtained transduced optical power at $\Omega_m$ ($P_1$) and the response obtained by Eq. \eqref{eqapprox2} are shown in (b) and (c) respectively. The plots are normalized to their respective peaks. The integrated RF spectral powers for each harmonic with increasing pump power are shown in (d). Similarly, for Probe 2, The sidebands are formed well outside the resonance placing it in the resolved-sideband regime as shown in (e). The corresponding theory and experimentally obtained response of the resonance are shown in (f) and (g) respectively, with the integrated RF spectra plotted in (h) for increasing pump power. $\Delta$: laser cavity detuning, $P_{in}$: pump input power, $P_{th}$: phonon lasing threshold power.}
  \label{fig4}
\end{figure*}

\section{Results}
\subsection{Nonlinear transduction}
In order to verify the proposed approach, we run phonon lasing measurements on a silicon microdisk resonator. The resonator has a radius of 4 $\mu$m, a thickness of 260 nm, and supports a radial breathing mechanical mode with a resonance frequency of 644 MHz and a quality factor of 8000 in vacuum as shown in Fig. \ref{fig2}(b) (see Appendix \ref{AppFab} for details on device fabrication). We run two pump-probe experiments with the same pump mode in both. In the first experiment we use a probe resonance, labeled as Probe 1 in Fig. \ref{fig2}(c), with a loaded linewidth of 3.42 GHz putting it in the unresolved-sideband regime with a ratio $\Omega_m/\Gamma_{t,1} =$ 0.19. For the second measurement we use Probe 2 as shown in Fig. \ref{fig2}(c) with a loaded linewidth of 157 MHz, putting it in the resolved-sideband regime with $\Omega_m/\Gamma_{t,2} =$ 4.1. Traditional wisdom would suggest that only Probe 1 would be able to measure modulation amplitudes larger than $\Gamma_{t,2}$. In these measurements, we will show that both resonances can measure modulation signals much larger than this. In order to excite the mechanical mode, a laser is coupled in via a tapered optical fiber brought into the near field of the resonator. The laser is blue detuned from the pump resonance center and the dropped power is increased. This causes heating of the mechanical mode and after a threshold, coherent mechanical oscillations occur \cite{lasersat1} with an energy that increases linearly with the pump power. A second laser monitors the motion of the two probe resonances with $\Delta=\Gamma_{t,1(2)}/2$ to measure these oscillations (see Appendix \ref{AppExp} for details on the experimental setup). The transmitted RF power at $\Omega_m$ for the 2 probe modes is shown in Fig. \ref{fig3}(a) and (d). As expected, Probe 2's response saturates preventing measurement of the cavity's motion beyond a certain modulation strength. In order to calibrate the amplitude of motion, we use Probe 2's average transmitted power when the probe laser scans through it. The strength of the sidebands formed around the resonance can be used to accurately calibrate the modulation amplitude. Appendix \ref{AppCalib} shows the details of the calibration. With this information we plot theory and experimental results for the two probes in Fig. \ref{fig4}. We set the zero of modulation amplitude at the lasing threshold power $P_{th}=10.8$ $\mu$W and plot probe response vs the square root of the pump input power $P_{in}$ as this is proportional to the amplitude of the resonator's motion. We see good agreement between theory and experimental results. A maximum modulation amplitude $g_{om}x_{0}=$ 952 MHz ($\eta\approx 1.5$) is obtained by a direct comparison of theory and experiment which is in agreement with the calibration done independently. This modulation amplitude is over six times larger than $\Gamma_{t,2}$. In spite of that, as Fig. \ref{fig4}(g) indicates, the response of the proposed approach does not saturate giving a correct measurement. Therefore the dynamic range for Probe 2 has been extended far beyond the saturation limit. We did not increase the modulation amplitude further as silicon suffers from strong two-photon absorption and this can cause broadening of the resonances corrupting the measurement. In Section \ref{secdisc}, we will theoretically show that our proposal works at modulation amplitudes much larger than what we have achieved experimentally. Figure \ref{fig4}(d) and (h) show the integrated electrical spectral power for the two probes measured at the three harmonics as they evolve with increasing mechanical motion. The spectral traces corresponding to each data point in Fig.  \ref{fig4} are plotted in  Fig. \ref{figS3}. An important point to note here is that in these measurements, we are not comparing sensitivities and dynamic ranges of the two probe modes to identify which regime of operation performs better. Rather, we have shown that the linewidth does not pose any limits to sensing with optical cavities and two drastically different modes can have linear response to modulation owing to a change in the detection bandwidth. This does not settle the question about the sensitivity-dynamic range trade-off. We will address this in the next section.

\begin{figure*}
  \includegraphics[scale=0.93]{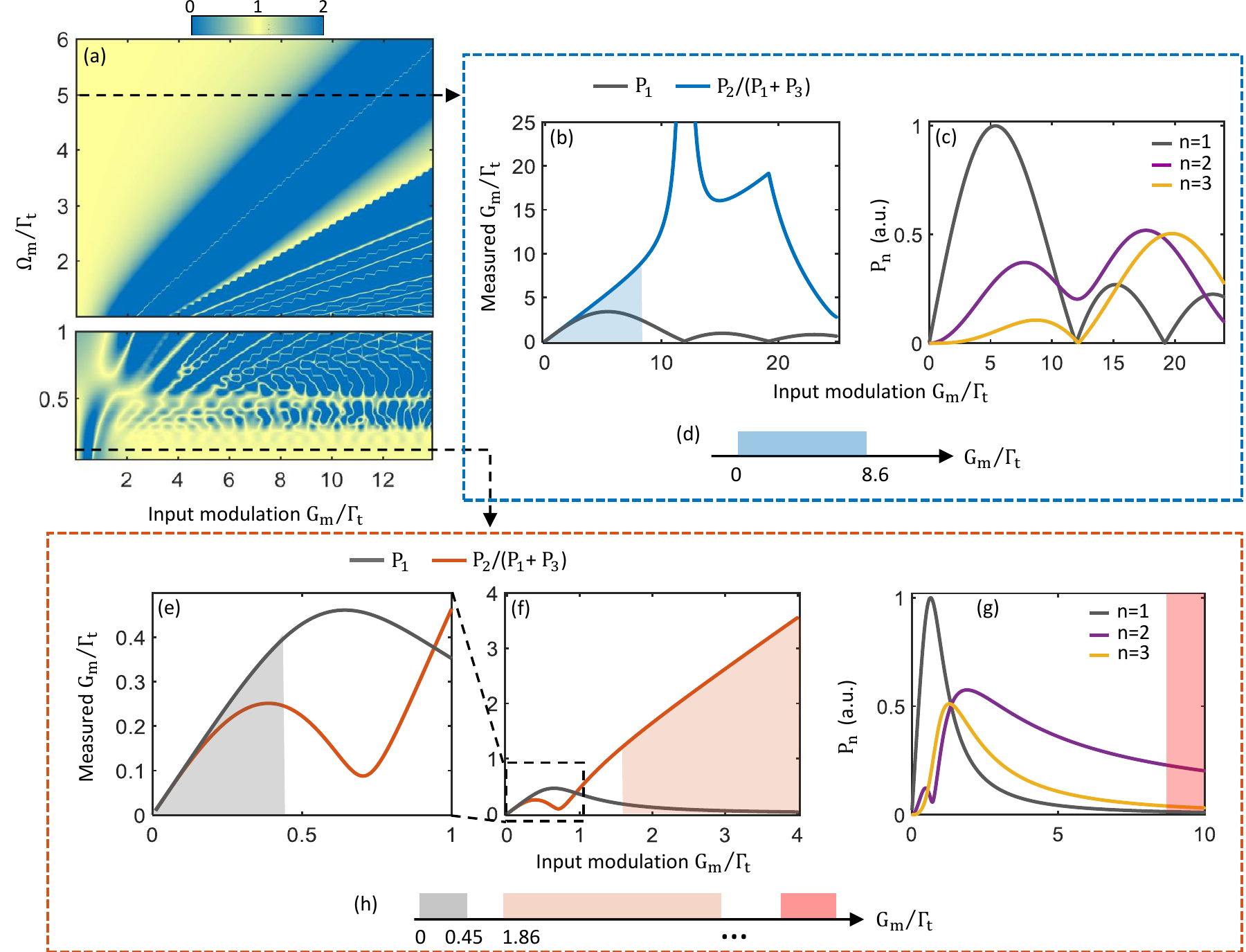}
  \centering
  \caption{Estimation of the upper limit of dynamic range. (a) The derivative of Eq. \eqref{eqamp} with respect to the modulation amplitude ($d/dG_m [P_2/(P_1+P_3)]$, $G_m=g_{om}x_0$) is numerically evaluated. The contour is normalized such that the small signal response has a unity slope. The regions with a constant positive contour are the areas where the response of the cavity is linear and can be used for sensing. (b) Using a slice of the contour at $\Omega_m/\Gamma_t=5$ in the resolved-sideband regime, the cavity's response is plotted. The legend indicates the two techniques used for measuring modulation and the harmonics involved in the measurement with the first harmonic only in gray and the proposed technique in blue, with the corresponding powers shown in (c). The shaded regions indicate the available dynamic range with the limit set by a 10\% deviation from linearity. Using these plots the achievable dynamic range is shown as colored sections on a $G_m/\Gamma_t$ axis in (d). A similar treatment is done in the unresolved-sideband regime at $\Omega_m/\Gamma_t=0.1$. (e) and (f) show the corresponding response of the cavity with the shaded regions identifying the dynamic range and the power in the three harmonies shown in (g). (h) shows the available dynamic range in this regime. The red shaded areas in (g) and (h) indicate the region where the powers in the first three harmonics decrease below the detection noise level.}
  \label{fig6}
\end{figure*}

\subsection{Dynamic range estimation} \label{secdisc}
The expressions providing the modulation amplitudes in Eqs. \eqref{eqamp} and \eqref{eqapprox2} were obtained under certain assumptions. In this section, we will investigate the limits to which these assumptions work, and show that the proposed approach works far beyond these limits. In the resolved-sideband regime, this limit is clear in Eq. \eqref{eqresol}, where the intra-cavity field is proportional to $J_0(\eta)$. This will vanish the intra-cavity power at modulation amplitudes $\eta$ where $J_0(\eta)=0$ breaking the approximation. Therefore the first zero of this Bessel function determines the true dynamic range of the proposed scheme. This is visualized in Fig. \ref{fig6}(a) where we have plotted the numerically calculated derivative of the ratio $P_2/(P_1+P_3)$ with respect to the modulation amplitude $G_m=g_{om}x_0$ in the transmitted power using the exact solution in Eq. \eqref{eqexact} for a critically coupled cavity mode. The dynamic range can then be seen as the areas in the plot where the contour is flat. Immediately, we can see such a region in the resolved-sideband regime. Setting $\Omega_m/\Gamma_t=5$, the cavity's response is plotted in Fig. \ref{fig6}(b). The linearity breaks when a singularity forms at $\eta=2.405$ ($G_m=12.03$) which is the first zero of $J_0(\eta)$ as expected. The available dynamic range is shaded in Fig. \ref{fig6}(b) using a 10\% deviation from linearity as the limit. This dynamic range is already larger than the saturation limit and will continue to increase as we increase the modulation frequency to go deeper into the resolved-sideband regime. This, however will come at the cost of sensitivity. This is because the transmitted power at the $n$-th harmonic is proportional to $J_n(g_{om} x_0/\Omega_m)$ and the slope of this function for small $x_0$ decreases as $\Omega_m$ increases. Simply put, the response of a cavity to a modulation decreases if the modulation frequency is made larger in the resolved sideband regime. Therefore, an arbitrarily large dynamic range can be obtained with at the cost of ever decreasing sensitivity and thus, keeping the trade-off intact.

\par Figure \ref{fig6}(a) shows another region with a constant contour in the unresolved-sideband regime. Setting $\Omega_m/\Gamma_t=0.1$ in this region, the cavity's response is plotted in Fig. \ref{fig6}(e)-(g). As mentioned in Section \ref{sectheory}, the assumption made was that the modulation signal changes adiabatically in time. We can see the assumption breaking as the modulation amplitude approaches $\Gamma_t/2$ as shown in Fig. \ref{fig6}(e). However, quite surprisingly, the assumption holds indefinitely after the modulation amplitude exceeds this value as seen in Fig. \ref{fig6}(f). Therefore Eq. \eqref{eqapprox2} provides an accurate measurement of arbitrarily large modulation amplitudes with a constant slope, thus breaking the sensitivity-dynamic range trade-off. 

\par When we consider a more realistic situation by including detector noise, there is an upper limit to Eq. \eqref{eqapprox2}. The noise-limited sensitivity of our proposed approach will be smaller than the conventional approach in which only the first harmonic is used and will be variable over the linear region of the response. This is due to the nonlinear transduction of each harmonic as seen in Fig. \ref{fig6}(g) and the fact that we need to make three measurements instead of one, which adds to the total noise. Appendix \ref{AppSen} shows a detailed comparison between the sensitivity of our approach with the conventional method. As the modulation amplitude increases, the optical power continues to shift to higher harmonics. Eventually, the power in the first three harmonics or the slope approach zero as seen in Fig. \ref{fig6}(g). Therefore, the true dynamic range in this regime will ultimately be limited by the detection noise level which will determine the minimum sensitivity that can be tolerated.

\par Given this information, the best approach will then be to combine our proposed technique with the conventional method, where the first harmonic is used for small modulation maintaining the small-signal sensitivity, and using the next two harmonics for large modulation amplitudes giving a large dynamic range. This way, the cavity can switch between two modes, taking best of the two worlds. For comparison, in order to meet the increased dynamic range with the conventional approach, the decrease in quality factor needed will cause significant loss in sensitivity (see Appendix Fig. \ref{figS5}) rendering the sensor useless for weak measurements. This demonstrates the advantage of an increased detection bandwidth in our approach.

\par Finally we can consider extension of our model to a multi-frequency excitation.  If the cavity is driven to oscillations with more than one frequency, then as long as the three harmonics of each frequency are distinguishable, the model will independently apply to each frequency and determine its modulation amplitude. If any of the harmonics overlap, then the model will fail. A situation where this will happen is when the optomechanical coupling has a quadratic component \cite{taylor2,supp-linear,supp-linear2}. Here the second harmonic generated due to the nonlinear transduction of the cavity will be indistinguishable from the linear transduction of the quadratic term causing our model to fail. Nonlinear optomechanical coupling can arise in some situations when the oscillator is driven at very high modulation amplitudes \cite{nanmechanical} putting an upper limit to our model specific to the physics of the oscillator. In the case of a purely quadratic coupling, our model will be linear in mechanical energy instead of the modulation amplitude and Eqs. \eqref{eqamp} and \eqref{eqapprox2} will provide the mechanical energy correctly in the nonlinear modulation regime.

\subsection{Phonon lasing saturation}
The investigation into the saturation limit of mechanical modulation motivated us to consider a separate but related issue of saturation in phonon lasing. After the lasing threshold is reached, the mechanical energy increases linearly with the input power. However, in most of the experiments done to date, this linearity does not persist and a saturation behavior is observed at higher powers. Saturation in phonon lasing can be due to many reasons including pump depletion \cite{lasersat1,lasersat2,lasersat3}, and various other nonlinear effects \cite{nonlinear-lasersat1,nonlinear-lasersat2,nonlinear-lasersat3,nonlinear-lasersat4}. However, due to the cavity's transduction of the mechanical oscillator being nonlinear as well, it becomes difficult to identify if the mechanical oscillation amplitude is in fact saturating. Given that our proposal linearizes the cavity transduction, we can differentiate between the two processes. For this purpose, we run another phonon lasing measurement with the Probe 2 resonance acting as both the pump and probe mode. Appendix \ref{Applasing} has details of the measurement and the calculation of the mechanical energy of the oscillator using our proposed approach.

\section{Conclusion}
In this article, we have investigated the upper limit of dynamic range for cavity optomechanical sensors. We have theoretically proposed and experimentally verified a technique that allows measurement of modulation amplitudes much larger than the cavity linewidth, extending the dynamic range into the nonlinear regime where the transduction of such large signals is ordinarily not possible. This is done by extending the detection bandwidth to measure the first three harmonics of the modulation signal. We further show theoretically that the dynamic range can be made arbitrarily large within a small detection bandwidth easing the trade-off between sensitivity and dynamic range that afflicts sensors across the board. It is interesting to note that saturation of a transducer to an overwhelmingly large input, a highly nonlinear process, can be rendered linear by a different choice of measurement due to the recurrence property of the Bessel function. This property has allowed us to accurately measure a signal even if most of its energy is present outside of the detection bandwidth, with only a fraction of the signal spectrum giving all the information. Although our experiment was done on a cavity optomechanical system, this approach can be applied to any cavity-based sensing system. We envision that this technique will help design cavity-based sensors that are simultaneously sensitive to small signals and allow for a large dynamic range to measure dynamic phenomena.

\begin{acknowledgments}
The authors would like to acknowledge support from National Science Foundation Grant No. EFMA-1641099, ECCS-1810169, and ECCS-1842691. Defense Threat Reduction Agency-Joint Science and Technology Office for Chemical and Biological Defense grant No. HDTRA11810047. This work was performed in part at the Cornell NanoScale Facility, a member of the National Nanotechnology Coordinated Infrastructure (NNCI), which is supported by the National Science Foundation (Grant NNCI-2025233).
\end{acknowledgments}

\bibliography{apssamp}

\appendix
\counterwithin{figure}{section}
\section{Deriving Eq. (5) from Eq. (3)} \label{AppEqn}
starting from Eq. \eqref{eqresol} in the main text, we have
\begin{align} 
    a(t)=-\frac{i\sqrt{\Gamma_{ex}}A_{in}J_0(\eta)}{i\Delta-\Gamma_t/2}\exp[{-i\eta \sin{(\Omega_m t)}}].
\end{align}
Plugging this into the expression for transmitted power
\begin{align}
    P(t)&=|A_{in}+i\sqrt{\Gamma_{ex}}a(t)|^2 ,\\ \nonumber
    &=|A_{in}|^2 \left|1+\frac{\Gamma_{ex}J_0(\eta)}{\sqrt{\Delta^2+(\Gamma_t/2)^2}}\exp{[-i(\phi_0 +\phi_m)]}\right|^2,\\ \nonumber
    &=|A_{in}|^{2} \Big\{ 1+\frac{\Gamma_{ex}^2J^{2}_{0}(\eta)}{\Delta^2+(\Gamma_t/2)^2}\\ \nonumber
    &+ \frac{2\Gamma_{ex}J_0(\eta)}{\sqrt{\Delta^2+(\Gamma_t/2)^2}}\cos{(\phi_m+\phi_0)}  \Big\},\\
    &= P_{DC}+\delta P(t),
 \end{align}
 where $\phi_0=\tan^{-1}(-2\Delta/\Gamma_t), \; \phi_m=\eta \sin{(\Omega_m t)},$ and $\delta P(t)$ is the time varying part of the transmitted power. Ignoring the DC component of the power, we have
 \begin{align}
    \delta P(t)=\frac{2|A_{in}|^2 \Gamma_{ex}J_0(\eta)}{\sqrt{\Delta^2+(\Gamma_t/2)^2}} [ \cos{(\phi_m)}\cos{(\phi_0)}\\ \nonumber
    -\sin{(\phi_m)}\sin{(\phi_0)} ].
 \end{align}
Setting $\Delta=\Gamma_t/2$, we get
\begin{align} \label{eqS1}
    \delta P(t)=\frac{2|A_{in}|^2 \Gamma_{ex}J_0(\eta)}{\Gamma_t} \left[ \cos{(\phi_m)}+\sin{(\phi_m)} \right].
 \end{align}
Using Bessel function properties
\begin{align}
    \cos{(\phi_m)}&=\cos{(\eta\sin{(\Omega_m t)})}\\ \nonumber
    &=J_{0}(\eta) +2\sum^{\infty}_{n=1} J_{2n}(\eta)\cos{(2n\Omega_m t)},
\end{align}
\begin{align}
    \sin{(\phi_m)}&=\sin{(\eta\sin{(\Omega_m t)})}\\ \nonumber
    &=2\sum^{\infty}_{n=0} J_{2n+1}(\eta)\sin{[(2n+1)\Omega_m t]}.
 \end{align}
Plugging this into Eq. \eqref{eqS1}, we get
\begin{align} 
   \delta P(t)&=\frac{2|A_{in}|^2 \Gamma_{ex}J_0(\eta)}{\Gamma_t} \Big\{ J_{0}(\eta) \\ \nonumber 
    &+2\sum^{\infty}_{n=1} J_{2n}(\eta)\cos{(2n\Omega_m t)} \\ \nonumber
    &+2\sum^{\infty}_{n=0} J_{2n+1}(\eta)\sin{[(2n+1)\Omega_m t]} \Big\},
 \end{align}
which is the Eq. (5).

\section{Device Fabrication} \label{AppFab}
The silicon microdisk resonator used for this experiment is fabricated on a standard Silicon-on-Insulator (SOI) wafer with a 260 nm thick single crystalline silicon on top of a 2$\mu$m thick oxide layer and silicon substrate. The microdisk is patterned using Electron-beam Lithography with a positive E-beam resist (ZEP520A). After resist development, the pattern is transferred from the resist layer to the silicon layer using an inductively-coupled plasma (ICP) - reactive-ion etching (RIE) process in a sulfer hexafluoride (SF6)/octafluorocyclobutane (C4F8) gas chemistry. Then the resist mask s removed using an oxygen plasma clean. Finally the microdisk is suspended by etching the oxide layer underneath using wet etching with Hydrofluoric acid (HF). To finely control the etching, the HF is diluted with water in a 1:10 ratio and etching is done in several steps of one to five minutes each. At each step, the resulting undercut is visually inspected using an optical microscope illuminated with polarized light and observed with an orthogonally oriented analyzer to reveal the edges of the structures (Polarized Light Microscopy). This is repeated until the microdisk pillar size is around 100-200 nm which is subsequently verified using a scanning electron microscope.

\section{Experimental Setup} \label{AppExp}
\begin{figure*}
  \includegraphics[scale=0.95]{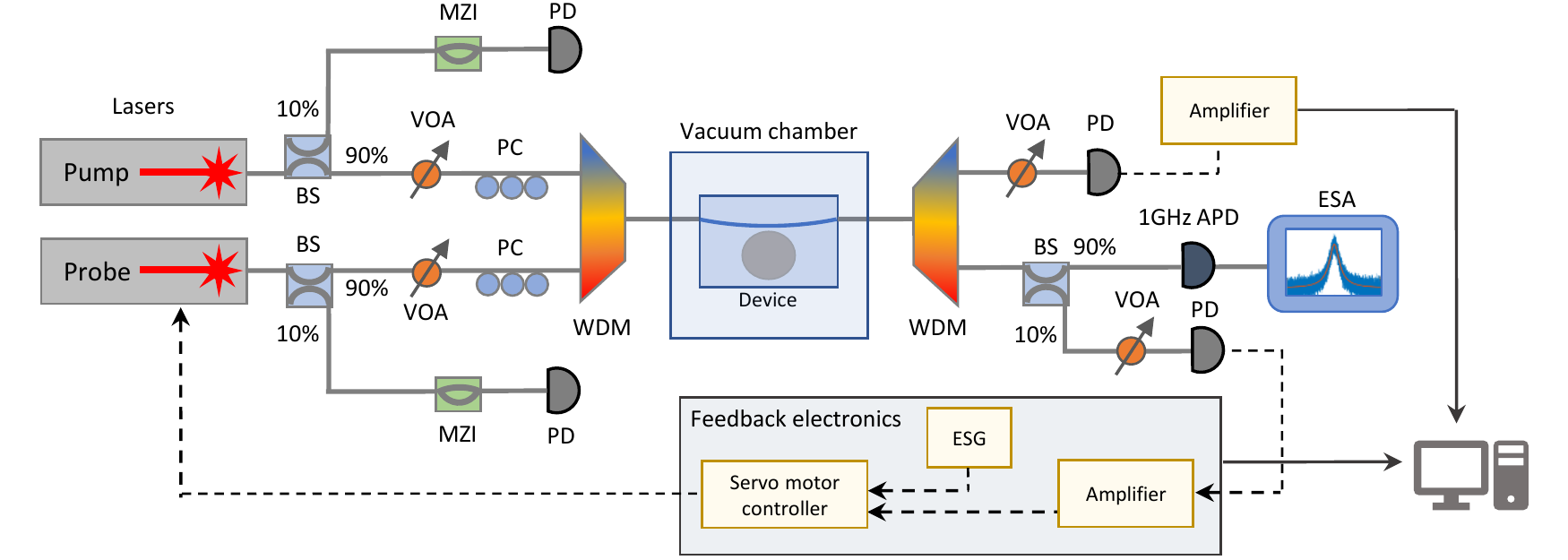}
  \centering
  \caption{Experimental setup for the pump-probe experiment. Two lasers are coupled into and out of the device using wavelength-division multiplexers (WDM). Only one laser is used for data in Appendix \ref{Applasing}.  The device is kept in a vacuum chamber at a pressure of $8\times10^{-4}$ mbar. The mechanical modulation is measured with a spectrum analyzer using an Avalanche photo-diode (APD). BS: Beam splitter, VOA: Variable optical attenuator, MZI: Mach-zehnder interferometer, PC: Polarization controller, PD: Photo-diode. ESA: Electronic spectrum analyzer, ESG: Electronic signal generator.}
  \label{figS6}
\end{figure*}

Figure \ref{figS6} shows the experimental setup. A strong pump laser is used to drive optomechanical oscillations in the microdisk resonator. The device is kept inside a vacuum chamber with a pressure of $8\times10^{-4}$ mbar. A weak probe laser measures the motion of the resonator using a second resonance (Probe 1 and Probe 2). The input power from the probe laser used to measure the mechanical motion was about 750 nW for Probe 1 and 370 nW for Probe 2. These powers are much smaller than the pump power which ranged from 6 $\mu$W to 28 $\mu$W and did not cause any competition between the lasers. Due to high optical power, the pump laser was thermally locked to the blue side of the pump resonance and did not show any measurable drift over the duration of the experiment. The probe laser was feedback locked using the fine wavelength adjustment piezo element inside the laser (New Focus TLB 6728). An aggressive feedback gain (New Focus LB1005-S) was used to correct for the fluctuations in the probe resonance frequency caused by the pump laser. A small fraction of the probe power (10\%) was routed into a photo-diode for this locking, while the rest of the power was sent to an Avalanche photo-diode (APD) and an RF spectrum analyzer. Both lasers were dropped into the resonator and separated out at the output using standard telecom-band wavelength-division multiplexers (WDM) with an isolation exceeding 100 dB.

\section{RF spectra for the pump-probe experiment} \label{AppSpec}
\begin{figure*}
  \includegraphics[scale=0.80]{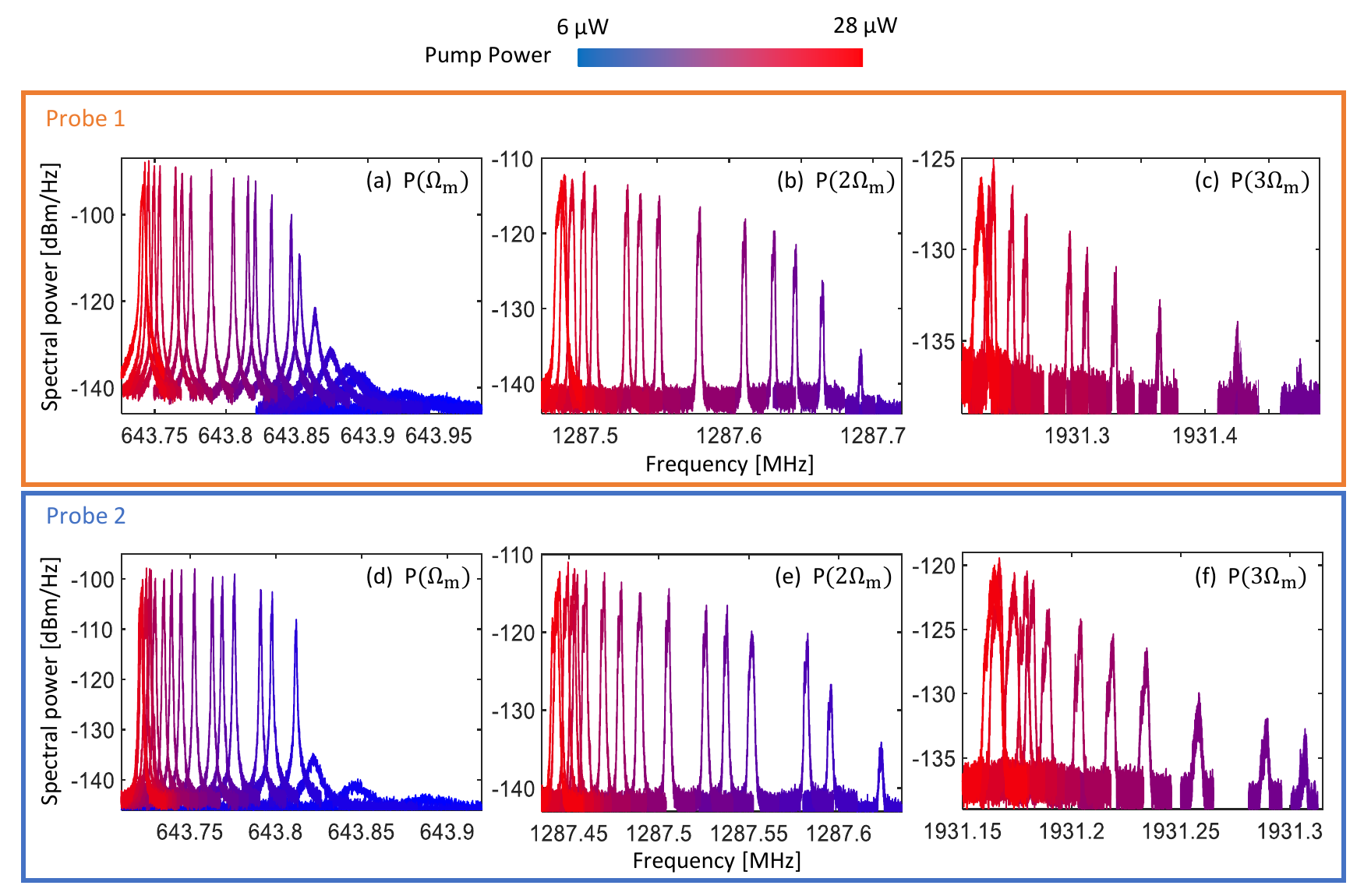}
  \centering
  \caption{RF spectra at the first three harmonics for the pump probe measurements in Fig. \ref{fig4}. The red shift in the spectra with increasing pump power is due to the optoemchanical back-action.}
  \label{figS3}
\end{figure*}

Figure \ref{figS3} Shows the measured RF spectra for the three harmonics of the Probe 1 and Probe 2 resonances. The data is collected with a 30-50 KHz integration window (depending on the spectral linewidth) and averaging over 50 spectra to obtain each plot. The shift of the spectra with increasing pump power is due to the optomechanical back-action that causes the mechanical resonance frequency to red-shift. The maximum shift in the center frequency of the spectra is only 0.03\% which is too small to cause any issues with our model. The power in each harmonic is determined by integrating the corresponding spectral trace.

\section{Callibration of modulation amplitude for Fig. \ref{fig4}} \label{AppCalib}

\begin{figure}
  \includegraphics[scale=0.90]{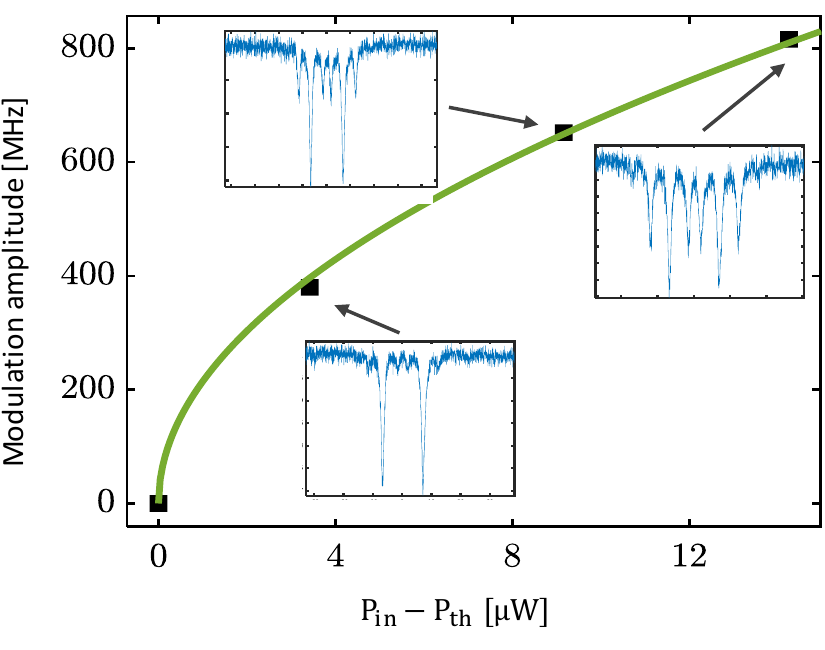}
  \centering
  \caption{Square root fitting of the modulation amplitude to the pump input power when the lasing threshold is set to zero. The insets show probe 2 resonance traces at three power levels indicated by arrows. Corresponding modulation amplitudes are obtained by the relative depth of the first sideband of the resonance to its central lobe.}
  \label{figS1}
\end{figure}

The ratio of average transmitted power at the first sideband of the resonance at $\Delta=\Omega_m$ to the depth of the resonance at $\Delta=0$ can uniquely determine the modulation amplitude for relatively small modulation where the sideband does not saturate. Using the exact solution of the modulated resonator in Eq. (2) in the main article, we have calculated the modulation amplitudes corresponding to average power transmission spectra obtained by scanning the laser across the probe 2 resonance for three values of the pump power as shown in Fig. \ref{figS1}. The data shows good agreement with a square root fit as seen in as the mechanical energy of the resonator is linear with the driving power in phonon lasing. Extrapolating the fitting, the modulation amplitude corresponding to the highest pump power i.e. the last data point in Fig. \ref{fig4}(g)  is 920 MHz. This is very close to 952 MHz, the modulation amplitude measured directly by comparing theory and experimentally obtained data for the optical power at $\Omega_m$ as plotted in Fig. \ref{fig4}(f) and (g). Therefore the  two methods for calibrating the mechanical motion have a decent agreement, with the small difference most likely due to the noise in the measured transmission spectra. We note that the normal mode splitting of the Probe 2 resonance is 1.72 GHz, much larger than the mechanical resonance frequency $\Omega_m$, and there is no overlap in the sidebands of the split modes that could complicate this measurement.

\section{Sensitivity of detection} \label{AppSen}

\begin{figure}
  \includegraphics[scale=0.95]{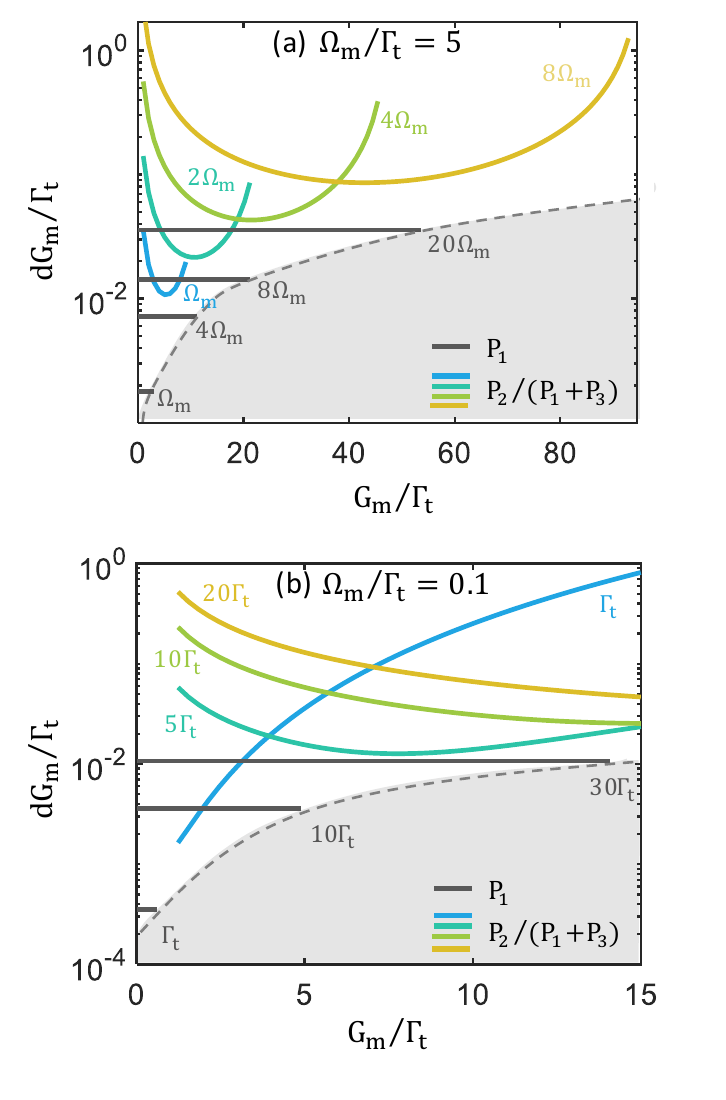}
  \centering
  \caption{(a) Sensitivity of detection for increasing modulation amplitude as calculated by Eqs. \eqref{eqsen1} and \eqref{eqsen123} for different modulation frequencies indicated with each curve. The lengths of each curve indicates the avaliable dynamic range using 10\% deviation from linearity with the straight lines in black evaluated using only the first harmonic. The gray shaded regions indicate the region of dynamic range inaccessible to the sensor. (b) same plots for the unresolved-sideband regime.}
  \label{figS5}
\end{figure}

To characterize the sensitivity of our proposed technique, we assume that the sensitivity is limited by the detection noise. We can evaluate the modulation sensitivity for the first harmonic by setting the signal-to-noise ratio SNR = 1 as
\begin{align} \label{eqsen1}
    \frac{dG_m}{G_m}=\frac{d P_1}{P_1}= \frac{\mathrm{NEP}}{P_1},
 \end{align}
where NEP is the noise-equivalent power, $G_m$ is the modulation amplitude and $dG_m$ is the corresponding sensitivity. Similarly for our proposed approach, the sensitivity can be evaluated to be 
\begin{align}\label{eqsen123}
    \left[\frac{dG_m}{G_m}\right]^2=\left[\frac{\mathrm{ NEP}}{P_2}\right]^2+2\left[\frac{\mathrm{NEP}}{P_1+P_3}\right]^2.
 \end{align}

Using an NEP of 1.6 pW/$\sqrt{Hz}$ (NewFocus 1647 APD), a 50 KHz integration bandwidth for each harmonic, and an operating power of 1 $\mu$W, we have evaluated the two sensitivities. The results are plotted in Fig. \ref{figS5}(a) for different cavity linewidths with $\Omega_m/\Gamma_t=0.1$ in the unresolved-sideband regime, and Fig. \ref{figS5}(b) for $\Omega_m/\Gamma_t=5$ with increasing modulation frequencies in the resolved-sideband regime. The length of the lines in the plot determines the dynamic range of sensing with the limit set by a 10\% deviation from linearity. Our proposed approach produces a variable sensitivity. The figures show that in order to increase the dynamic range, orders of magnitude in sensitivity will be lost if only the first harmonic is used as indicated by the gray lines. In contrast this sensitivity for weak perturbations can be maintained with our proposed technique. In addition, our technique even beats this sensitivity for a part of the dynamic range in the unresolved sideband regime as seen in the blue curve (linewidth = $\Gamma_t$) in Fig. \ref{figS5}(b) . The true dynamic range (limited by detector's noise) can be several times larger than this depending on how much loss of sensitivity can be tolerated. Therefore, low noise detectors, with high responsivity, operated at high optical powers will help increase the workable dynamic range.

\section{Saturation effects in phonon lasing experiments} \label{Applasing}
\begin{figure}
  \includegraphics[scale=0.9]{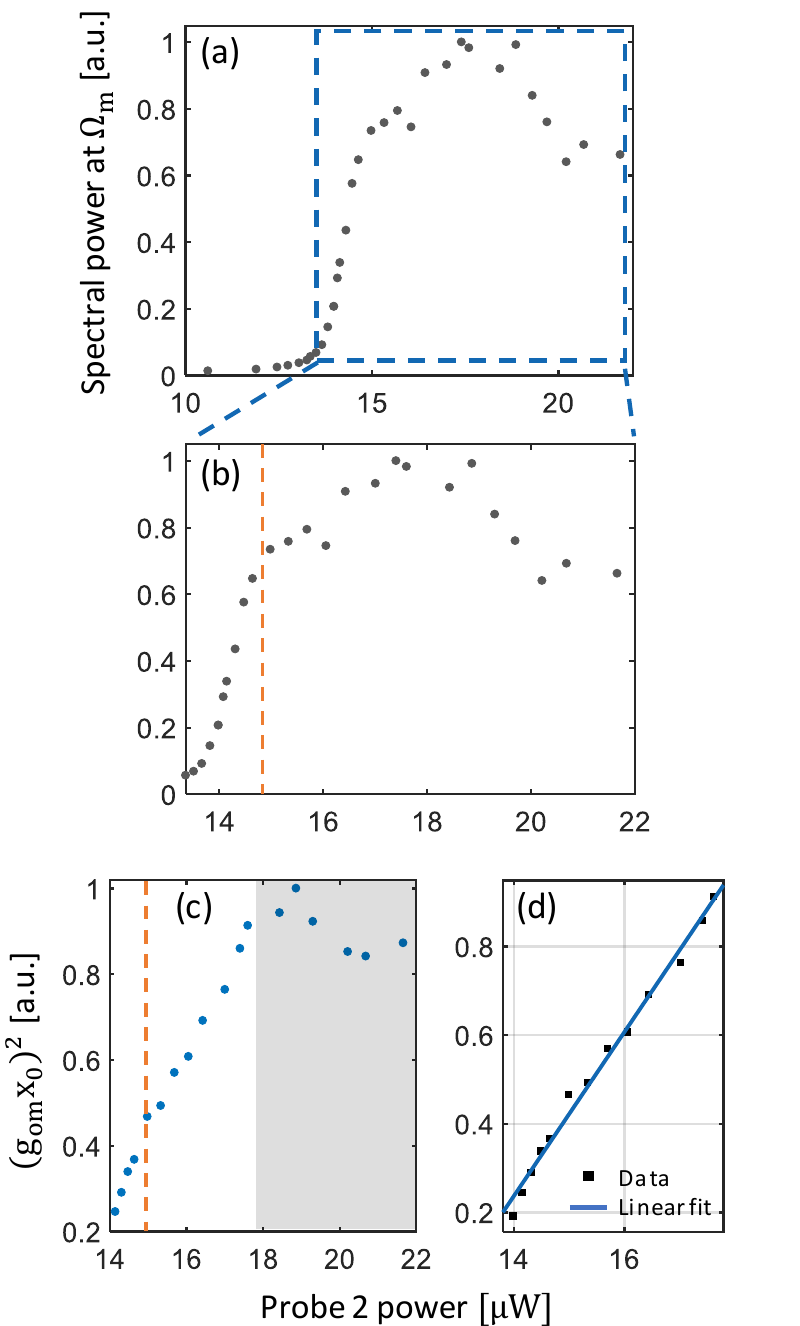}
  \centering
  \caption{Phonon lasing with the Probe 2 mode. (a) Mechanical energy of the resonator obtained by integrating the RF spectral power at the first harmonic only. (b) A section of the plot in (a) after the lasing threshold at an input power of 13.8 $\mu$W is reached showing saturation of the energy around 15 $\mu$W indicated by a dashed line. (c) Mechanical energy calculated by squaring Eq. (7) (main article) with the shaded region indicating deviation from linearity. (d) Fitting of the lienar part of the data in (c) to a line.}
  \label{fig5}
\end{figure}
\begin{figure*}
  \includegraphics[scale=0.9]{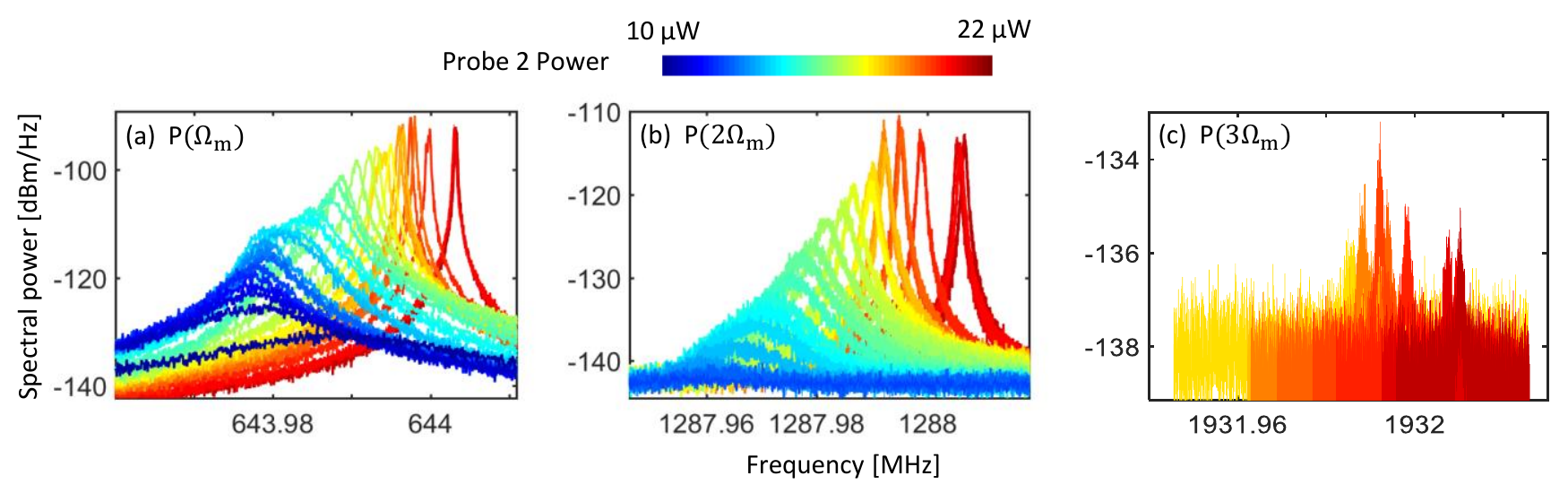}
  \centering
  \caption{RF spectra at the first three harmonics for the phonon lasing measurements in Fig. \ref{fig5}.}
  \label{figS4}
\end{figure*}

As mentioned in the main text, saturation in phonon lasing can be due to various nonlinear effects including pump depletion. However, the nonlinear transduction of the resonacne also causes saturation of the transduced signal. We will differentiate between the two processes using our proposed approach of measuring the first three harmonics of the lasing frequency. Probe 2 acts as both pump and probe mode as is usually done in these measurements. The mode is coupled more strongly to the tapered fiber to increase the linewidth so that the laser can be locked more easily. In this measurement the Probe 2 linewidth is 258 MHz and $\Omega_m/\Gamma_{t,2}$ = 2.5. The laser is blue detuned to the mechanical sideband and the transmitted power from Probe 2 itself is used to measure the mechanical motion instead of a second resonance acting as a probe, as this is how a lasing measurement is usually done. Figure \ref{fig5} shows the results of the measurement. Here we have plotted the integrated electrical spectrum as this gives the mechanical energy of the oscillator. As mentioned earlier, the transmitted power at the first harmonic initially grows linearly, after the lasing threshold is achieved, but soon saturates. The mechanical energy obtained by squaring Eq. (7) is plotted in Fig. \ref{fig5}(c). The data indicates that the mechanical oscillator saturates when the input power is around 18 $\mu$W as indicated by the shaded region on the plot as opposed to the saturation that occurs around 15 $\mu$W in the first harmonic power as indicated by a dashed line in Fig. \ref{fig5}(b). This suggests that the mechanical mode continues to heat up after this power and the nonlinear transduction of the cavity is the cause of the early saturation of the first harmonic signal. At even higher input power, the mechanical oscillator starts to cool down. The saturation and subsequent cooling of the oscillations are likely due to a combination of pump depletion and the effects of two-photon absorption causing a saturation of intra-cavity power and broadening the resonance shifting the laser detuning. A linear fitting of the data is shown in Fig. \ref{fig5}(d). Even though the theory does not strictly apply in this case as the optical mode measuring the motion of the resonator is also driving the mechanics, the data fits well with a straight line. The RF spectra corresponding to each data point in Fig. \ref{fig5} are plotted in Fig. \ref{figS4}.

\end{document}